\documentclass[10pt,conference]{IEEEtran}

\usepackage{alltt                                    
          , multirow
          , booktabs
          , listings
          , graphicx
          ,float
	,cite
          ,verbatim
         ,mathtools
	,url
	,amsmath
}
\usepackage[table]{xcolor}
\usepackage[numbers]{natbib}     
\usepackage{syntax}
\usepackage{algorithmic, algorithm}
\usepackage{enumitem}
\usepackage{threeparttable}
\usepackage{framed}
\usepackage{hhline}
\usepackage{balance}

\usepackage{expl3}
\ExplSyntaxOn
\newcommand\latinabbrev[1]{
  \peek_meaning:NTF . {
    #1\@}%
  { \peek_catcode:NTF a {
      #1., \@ }%
    {#1., \@}}}
\ExplSyntaxOff

\newcommand{\CASE}[1]{\STATE \textbf{case} #1\textbf{:} \begin{ALC@g}}
\newcommand{\ENDCASE}{\end{ALC@g}}

\newcommand{\DEFAULT}{\STATE \textbf{default:} \begin{ALC@g}}
\newcommand{\ENDDEFAULT}{\end{ALC@g}}
\newcommand{\DEFAULTLINE}[1]{\STATE \textbf{default:} }
\let\footnotesize\scriptsize

\newsavebox{\supbox}
\newcommand{\bsup}{\begin{lrbox}{\supbox}$\tt\scriptstyle}
\newcommand{\esup}{$\end{lrbox}{}^{\usebox{\supbox}}}
\def\eg{\latinabbrev{e.g}}
\def\ie{\latinabbrev{i.e}}

\definecolor{lightpurple}{rgb}{0.8,0.8,1}
\definecolor{codebg}{RGB}{255,255,255}
\definecolor{commentcolor}{RGB}{11,140,11}
\begin{document}
%

\title{Impact of Continuous Integration on Code Reviews\vspace{-.3cm}}
%
%
%
%
%

\author{\IEEEauthorblockN{Mohammad Masudur Rahman  ~~~ Chanchal K. Roy}
\IEEEauthorblockA{Department of Computer Science, University of Saskatchewan, Canada\\
\{masud.rahman, chanchal.roy\}@usask.ca}
}

\maketitle

\begin{abstract}
Peer code review and continuous integration often interleave with each other
in the modern software quality management.
Although several studies investigate how non-technical factors (\eg\ reviewer workload), developer participation and even patch size affect the code review process, 
the impact of continuous integration on code reviews is not yet properly understood.
In this paper, we report an exploratory study using 578K automated build entries where we investigate 
the impact of automated builds on the code reviews.
Our investigation suggests that successfully passed builds are more likely to encourage new code review participation in a pull request.
Frequently built projects are found to be maintaining a steady level of reviewing activities over the years, which was quite missing from the rarely built projects. 
Experiments with 26,516 automated build entries reported that our proposed model can identify 64\% of the builds that triggered new code reviews later.


\end{abstract}


\IEEEpeerreviewmaketitle

\section{Introduction}
\begin{figure}[!t]
\centering
\includegraphics[width=2in]{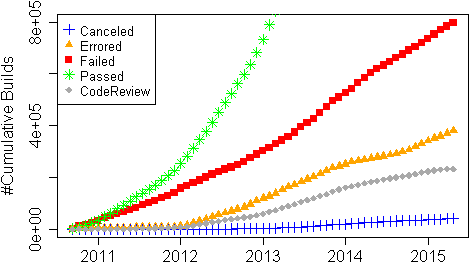}
\vspace{-.3cm}
\caption{Automated software build status and code reviews over the years}
\vspace{-.6cm}
\label{fig:motiv}
\end{figure}

Quality assurance is one of the most important steps of software change management which is often done using a combination of manual (\ie\ code reviews) and automated processes (\ie\ continuous integration).
GitHub, one of the most popular software ecosystems, offers a set of features for software quality management through pull requests (\ie\ for code reviews) and automatic build supports (\eg\ collaboration with Travis-CI).  
While peer code reviews involve manual checking of coding standard violations and simple logical errors in a submitted patch by the developers, 
continuous integration prevents the regression of a system by ensuring that the patch passes all the unit tests and merges with the system successfully.
In the pull-based modern software development, automatic builds are often 
interleaved with or followed by code reviews by the developers \cite{travis}.
Fig. \ref{fig:motiv} shows a growing trend for adopting automated software builds and code reviews by 1000+ open source projects from GitHub over the last six years. 
Peer code reviews are reported to be effective for improving coding standards \cite{panichella}, design quality \cite{fotuse-design} and overall quality \cite{shane-quality} of a software system.
There have been also several studies on how non-technical factors \cite{nontechnical}, developer participation \cite{olga-participation} and even patch size \cite{contempo,useful} affect the code reviews. 
Unfortunately, the impact of continuous integration on code review process is not yet properly understood given that
they are interleaving steps in the software quality management.
 In particular, how automated software builds and tests might influence the participation or overall quality of code reviews is not substantially studied.

In this paper, we report an exploratory study where we analyze the recorded logs of the thousands of automated builds performed on the open source projects at GitHub, and find out how they might affect the peer code review activities on the same projects. 
In particular, we investigate whether the status and frequency of the automated builds correlate to the participation or quality of the code reviews. Such findings are likely to help us better understand how the manual code review process could be complemented with automatic tool supports.
We thus answer three research questions as follows:
\begin{itemize}
\item \textbf{RQ$_1$:} Does the status of automated builds influence the code review participation in open source projects?
\item \textbf{RQ$_2$:} Do frequent automated builds help improve the overall quality of peer code reviews?   
\item \textbf{RQ$_3$:} Can we automatically predict whether an automated build would trigger new code reviews or not?
\end{itemize}

Exploratory study using 578K automated build entries from 1000+ open source projects suggested important correlations between continuous integration and code review activities. 
First, 
automated build status has a notable impact on code review participation in the projects.
Passed builds are more likely to trigger new code reviews for a pull request than the other builds.
Second, automated build frequency has a major role in improving code reviews of the open source projects. 
Our investigation suggests that frequently built projects are likely to maintain a steady level of reviewing activities over the years, which was quite missing from the rarely built projects. Experiment using 26,516 build entries reported that our model can identify 64\% of the builds that triggered new code reviews later, which is promising. Our model can offer automatic supports in the code reviews and software quality management by identifying the appropriate pull requests for code reviews.

\section{Data Collection}
\label{sec:data}
We collect a total of 578K automated build entries from MSR challenge dataset \cite{travis} for our study. 
Since we are interested to investigate possible relationships between automated builds and peer code reviews,
the collected entries should be associated with code reviews.
In GitHub, pull requests are generally used for code reviews.
Hence, we collect only such entries where each of the corresponding builds was triggered by a pull request (\ie\ \texttt{gh_is_pr$=$true}) submitted by the developer. We also identify whether the commits in each build underwent (\ie\ \texttt{gh_num_pr_comments$>$0}) and did not undergo (\ie\ \texttt{gh_num_pr_comments$=$0}) peer code reviews. 
We found 40\% (\ie\ 232K) of such commits reviewed by the developers. 
We extract automated build and test details (\eg\ outcome, frequency) and code review activities (\eg\ review comment statistics) from each of the collected entries for our comparative and inferential analysis.  
Table \ref{table:ds} shows the details of our collected dataset for the study.

\begin{table}
\centering
\caption{Dataset for Exploratory Study}\label{table:ds}
\vspace{-.2cm}
\resizebox{3.5in}{!}{%
\begin{threeparttable}
\begin{tabular}{l|c|c||c|c||c|c}
\hline
\multirow{2}{*}{\textbf{Build Status}} & \multicolumn{2}{c||}{\textbf{Automated Build Only}} & \multicolumn{2}{c||}{\textbf{Build + Code Review}} & \multicolumn{2}{c}{\textbf{Total}}\\
\hhline{~------}
& \textbf{Entry} & \textbf{Project} & \textbf{Entry} & \textbf{Project} & \textbf{Entry} & \textbf{Project}\\
\hline
\hline
 Canceled & 2,616 & 135  & 1,368  & 85 & 3,984 & 207\\
\hline
Errored & 51,729 & 2,138  & 27,262 & 1,673 & 78,991  & 2,735 \\
\hline
Failed & 55,546 & 2,368 & 39,025 & 2,139 & 94,571  & 3,106 \\
\hline
Passed & 236,573 & 5,774 & 164,174 & 5,299& 400,747  & 7,319 \\
\hline
\end{tabular}
\centering
\end{threeparttable}
}
\vspace{-.6cm}
\end{table}

\vspace{-.1cm}
\section{Answering RQ$_1$: Automated Build Status and Code Review Participation}
\label{sec:exploratory}
To answer RQ$_1$, we divide automated build entries into two non-overlapping groups-- builds with code reviews and builds without code reviews. 
The goal is to contrast between these two groups in terms of their build status and code review activities. 
An automated build can have one of these four statuses-- \emph{canceled}, \emph{errored}, \emph{failed} and \emph{passed}.
On the other hand, review comment counts from each entry could be considered as a proxy to code review participation. 
That is, if such a comment count is greater than zero, the commits associated with the build entry received code reviews and vice versa.
Alternatively, one or more reviewers participated and the participation is denoted as ``1" and vice versa. 
In GitHub, reviewers can submit two types of code review comments--\emph{ in-line comments} and \emph{summary comments} -- where they are also called as \emph{pull request comments} and \emph{issue comments} respectively.
Thus, we consider one independent variable (\ie\ automated build status) and two response variables (\ie\ pull request comment count and issue comment count) for RQ$_1$, and perform statistical tests and further analyses to answer the research question 
in terms of a hypothesis as follows:
\OuterFrameSep0pt
\FrameSep2pt
\begin{framed}
\noindent
\textbf{H1$_{0}$}: Code review participation is not affected by the status of previous automated builds.

\noindent
\textbf{H1$_{a}$}: Code review participation is significantly affected by the status of previous automated builds.
\end{framed}
\textbf{Test of Hypothesis:} While 40\% of the built commits from our dataset received one or more code reviews, developers did not participate in the code reviews of rest 60\% commits from the same set of projects. Please note that 95\%--100\% of the projects from the MSR challenge dataset \cite{travis} adopt pull request based code reviews. Therefore, such lack of reviews might not be a mere coincidence and thus warrants an in-depth analysis.
To ensure a fair comparative analysis, we pick up a random sample of 231,829 entries without code reviews (\ie\ equal to the entries with reviews) from our dataset.
Then we perform \emph{Chi-squared tests} on the samples combining both groups, and investigate whether the code review participation is independent of corresponding build status or not. The test reported a \emph{p-value} of 2.2e-16$<$0.05, which refutes the null hypothesis (H1$_0$). 
That is, alternative hypothesis (H1$_a$) is accepted, and code review participation is significantly affected by the status of automated builds.

\begin{figure}[!t]
\centering
\includegraphics[width=3.1in]{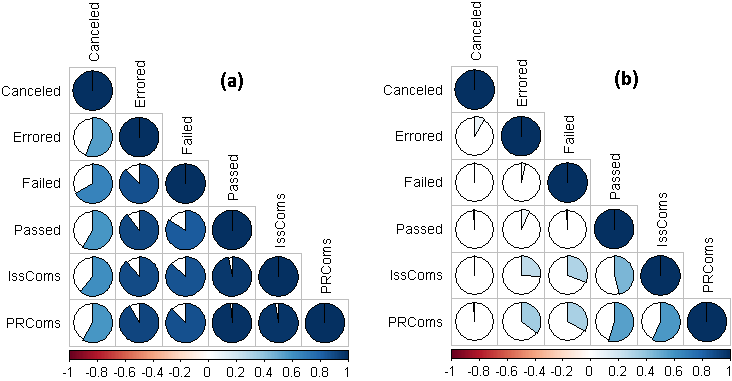}
\vspace{-.3cm}
\caption{Pearson correlation between build status and code review participation associated with (a) projects and (b) pull requests. IssComs=Issue comments, PRComs=PR comments}
\vspace{-.5cm}
\label{fig:corr-stat-rev}
\end{figure}

\textbf{Correlation Analysis:} Although the alternative hypothesis (H1$_a$) is found true according to the above statistical test, we perform correlation analysis between independent and response variables for gaining further insights. In particular, we consider two entities with different abstraction levels--\emph{project} and \emph{pull request}, and calculate their build statistics and code review statistics.     
We calculate the number of \emph{canceled}, \emph{errored}, \emph{failed} and \emph{passed} automated builds for each entity while determining the frequency of their built commits that received code reviews later.
The goal is to find out whether certain build status is correlated to code review participation or not.
Fig. \ref{fig:corr-stat-rev} shows the correlation plot between four build statuses and code review participation for (a) projects and (b) pull requests. 
We see that commits with \emph{passed} builds  received the maximum code reviews both in project level and in pull request level.
That is, reviewers possibly get more confidence in reviewing such code that is syntactically correct and meets functional requirements, \ie\ passes the automated builds and tests.
However, as shown in Fig. \ref{fig:corr-stat-rev}, the \emph{errored} and \emph{failed} builds also triggered notable review participation.

\begin{table}
\centering
\caption{Pull Requests with Code Review Comments Changed}\label{table:rev-change}
\vspace{-.2cm}
\resizebox{3.5in}{!}{%
\begin{threeparttable}
\begin{tabular}{l|c|c||c|c||c|c|c}
\hline
\textbf{Previous} & \multicolumn{2}{c||}{\textbf{Issue Comments}} & \multicolumn{2}{c||}{\textbf{PR Comments}} & \multicolumn{3}{c}{\textbf{All Review Comments}}\\
\hhline{~-------}
\textbf{Build Status}& \textbf{Add$\uparrow$} & \textbf{Remove$\downarrow$} & \textbf{Add$\uparrow$} & \textbf{Remove$\downarrow$} & \textbf{Add$\uparrow$} & \textbf{Remove$\downarrow$} & \textbf{Total}$\uparrow\downarrow$ \\
\hline
\hline
 Canceled & 15 & 21  & 9  & 7 & 20 & 24 & 65\\
\hline
Errored & 448 & 198  & 232 & 122 & \textbf{510}  & \textbf{265} & \textbf{812} \\
\hline
Failed & 1,379 & 676 & 610 & 299 & \textbf{1,542}  & \textbf{826} & \textbf{2,316} \\
\hline
Passed & 3,711 & 1,388 & 2,048 & 791 & \textbf{4,235}  & \textbf{1,788} & \textbf{5,677} \\
\hline
\end{tabular}
$\mathbf{\uparrow}$ = One or more review comments added, $\mathbf{\downarrow}$ = One or more comments removed from the reviews
\centering
\end{threeparttable}
}
\vspace{-.6cm}
\end{table}

\textbf{Review Change Analysis:} Review comment statistics for each entry in the challenge dataset are generally calculated for the time gap between the previous build and the current build on the same pull request \cite{travis}. That means, if such statistics are found changed during the current build submission, the previous build might have played a role given that corresponding logs reported the details of automated builds and tests performed.
We thus analyze the automated build sequence (\ie\ based on build starting time) of each pull request from our dataset, and determine their review comment changes. In particular, we identify the status of the immediate previous build and the change direction of review statistics in the current build for each of the pull requests.
Table \ref{table:rev-change} shows the statistics
 of pull requests that underwent such review changes.
We see that 28\% (8,870) of 31,648 pull requests (\ie\ associated with code reviews) received further code reviews which might have triggered by the immediate previous builds.
Passed builds are associated with most of these review changes (\ie\ 18\%) which confirms our findings from the correlation analysis.
Besides, as shown in Table \ref{table:rev-change}, failed and errored builds also introduced a moderate amount (\ie\ about 10\%) of reviews. 
We also check three file change statistics for each of these pull requests from the dataset, and 
found that 99\%--100\% of them underwent file changes. That is, automated builds first trigger the code changes which in turn might warrant further code reviews.


Thus, to answer \textbf{RQ$_1$}, automated build statuses are very likely to affect code review participation.
Passed builds encourage more code reviews than failed or errored builds.
While a code review process is generally initiated by the patch submitter requesting peers for reviews, 
automated builds along with their outcomes might also trigger further code reviews. 

\section{Answering RQ$_2$: Automated Build Frequency and Code Review Quality}
Our analyses in RQ$_1$ suggest that automated build status might affect code review participation while \emph{passed} builds having the maximum influence. Our conjecture is that developers possibly felt comfortable in reviewing such code (\ie\ investing their effort into) that has already gained a quality threshold (\ie\ passed the automated builds and tests).   
However, too many and too frequent builds might introduce reviewing job overload \cite{nontechnical}  
on the developers which is likely to hurt the review quality.   
We consider review comment counts as a \emph{proxy} to code review quality \cite{useful,devsee}. 
Thus, an investigation is warranted on how frequency of automated builds might affect the overall code review quality. 
We collect project level build statistics and review comment statistics, and contrast between the projects with high frequency builds and the projects with low frequency builds. 
We performed statistical tests and further analyses to answer the research question in terms of a hypothesis as follows:
\OuterFrameSep0pt
\FrameSep3pt
\begin{framed}
\noindent
\textbf{H2$_0$}: Quality of code reviews associated with highly frequent builds is similar to that of less frequent builds.

\noindent
\textbf{H2$_a$}: Quality of code reviews associated with highly frequent builds is significantly higher than that of less frequent automated builds.
\end{framed}

\begin{table}
\centering
\caption{Code Review Comments of Two Quartiles}\label{table:quartile}
\vspace{-.2cm}
\resizebox{3.5in}{!}{%
\begin{threeparttable}
\begin{tabular}{l|c|c|c||c|c|c||c|c|c}
\hline
\multirow{2}{*}{\textbf{Quartile}} & \multicolumn{3}{c||}{\textbf{Issue Comments}} & \multicolumn{3}{c||}{\textbf{PR Comments}} & \multicolumn{3}{c}{\textbf{All Review Comments}}\\
\hhline{~---------}
& \textbf{Mean} & \textbf{p-value} & $\mathbf{\Delta}$ & \textbf{Mean} & \textbf{p-value} & $\mathbf{\Delta}$ & \textbf{Mean} & \textbf{p-value} & $\mathbf{\Delta}$\\

\hline
\hline
 \textbf{Q$_1$}  & 0.60 & \multirow{2}{*}{$<$0.001*}& \multirow{2}{*}{0.35}  & 0.24  & \multirow{2}{*}{$<$0.001*} & \multirow{2}{*}{\textbf{0.49}} & 0.84 & \multirow{2}{*}{$<$0.001*} & \multirow{2}{*}{\textbf{0.41}} \\
\hhline{--~~-~~-~~}
\textbf{Q$_4$}  & 0.99 &  &   & 0.52 &  &   & 1.50 & &  \\
\hline
\hline

\textbf{Q$_1'$}  & 0.62  & \multirow{2}{*}{$<$0.001*}&  \multirow{2}{*}{\textbf{0.40}}  &  0.32  & \multirow{2}{*}{$<$0.001*} &  \multirow{2}{*}{\textbf{0.53}} & 0.94 & \multirow{2}{*}{$<$0.001*} & \multirow{2}{*}{\textbf{0.44}} \\
\hhline{--~~-~~-~~}
\textbf{Q$_4'$}  & 0.97 &  &   &   0.54 &  &   & 1.51 & &  \\
\hline
\end{tabular}
\textbf{*} = Statistically significant, $\mathbf{\Delta}$ = Cliff's delta for effect size, \textbf{Q$_i'$} = Quartile for total build counts
\centering
\end{threeparttable}
}
\vspace{-.6cm}
\end{table}

\begin{figure}[!t]
\centering
\includegraphics[width=2.2in]{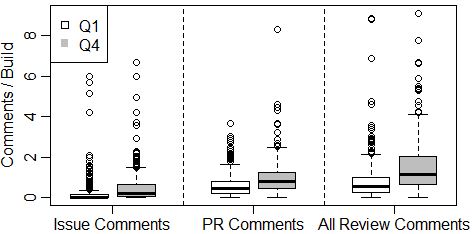}
\vspace{-.2cm}
\caption{Code review quality of projects from two different build frequency quartiles }
\vspace{-.5cm}
\label{fig:quartile}
\end{figure}

\textbf{Test of Hypothesis}: We calculate build frequency per month for each of the projects in our dataset.
Then we divide them into four quartiles where first quartile (Q$_1$) contains the lowest 25\% and fourth quartile (Q$_4$) contains the highest 25\% of all the build frequencies.   
We collect the corresponding project entries from both quartiles, and compare their mean review comment counts-- \emph{issue comment counts/build} and \emph{pull request comment counts/build}. 
We performed \emph{Mann-Whitney Wilcoxon} tests, and found that their review comment counts differ significantly, \ie\ all \emph{p-values} are less than 0.05. Table \ref{table:quartile} reports the details of our statistical tests, and Fig. \ref{fig:quartile} shows the box plots of mean comment counts for the two quartiles.
We see that code review quality (\ie\ mean review comment count) is significantly higher for Q$_4$ (\ie\ projects with frequent builds) than Q$_1$ (\ie\ projects with less frequent builds) in terms of all measures-- \emph{mean}, \emph{p-value} and \emph{Cliff's delta (effect size)}--
which refutes the null hypothesis (H2$_0$).
We also repeat the same experiments by considering total build counts rather than frequency per month for each of the projects, and reached at the same conclusion. Thus, build frequency has a significant impact on the code review quality of the projects.
Since automated tests are performed simultaneously with automated builds and thus have the same frequencies, the above findings regarding code review quality also equally apply to them.

\begin{figure}[!t]
\centering
\includegraphics[width=2.1in]{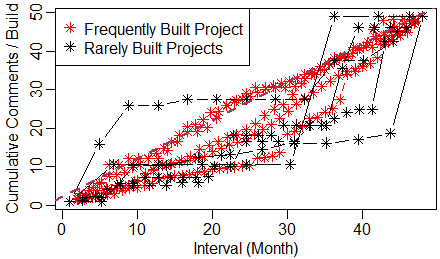}
\vspace{-.3cm}
\caption{Review comments of frequently built projects and rarely built projects}
\vspace{-.8cm}
\label{fig:rev-compare}
\end{figure}

\textbf{Interval-Aware Comparative Analysis:} While the above statistical analysis shows that the alternative hypothesis (H2$_a$) is true, we performed further analysis to gain more insights. In particular, we examine and compare the code review statistics of frequently built projects (\ie\ fourth quartile, Q$_4$) and rarely built projects (first quartile, Q$_1$) over specific time interval such as \emph{month}.    
We select Top-5 projects from Q$_4$ and Top-5 projects from Q$_1$ where each of the projects is 3--4 years old.
Then, we calculate review comments/build for each of the 48 months for the projects, and plot the cumulative comments/build for each project.
From Fig. \ref{fig:rev-compare}, we see that the cumulative curves for the frequently built projects are close to linear with an upward slope.
This suggests that these projects maintained steady review activities over the time period in question.  
On the other hand, the curves for the less frequently built projects are zigzag and do not show a regular structure. 
That is, they failed to maintain the regular code review activities, which is also manifested by their declining review comment statistics.   

%

Thus, to answer \textbf{RQ$_2$}, automated build frequency is very likely to have an impact on the quality of code reviews in the open source projects. 
While frequent builds help maintain an acceptable code quality standard for the projects, they also 
help trigger more code reviews than the infrequent builds.

\section{Answering RQ$_3$: Automatic Prediction of New Code Reviews using a Model}\label{sec:model}
Automated builds and tests often introduce source code changes which in turn might or might not trigger further code reviews.  
According to our analysis, about 100\% of the builds from the dataset are associated with code level changes. However, code reviews were triggered for only 40\% of the pull requests (details in Section \ref{sec:data}).  
Automated identification of such pull requests and their builds beforehand could help the project developers or stake holders with important decision making on code reviews.
Thus, a prediction model is warranted which can automatically predict whether an automated build is likely to trigger new code reviews or not.

We develop a prediction model where the model is trained on build log data using three machine learning algorithms.   
In particular, we use automated build status-- \texttt{tr_status}, three code change statistics--\texttt{gh_diff_files_added}, \texttt{gh_diff_files_deleted} and \texttt{gh_diff_files_modified}, two test change statistics-- \texttt{gh_diff_tests_added} and \texttt{gh_diff_tests_deleted} 
and two review comment statistics for the pull request--\texttt{gh_num_issue_comments} and \texttt{gh_num_pr_comments} -- as the predictor variables.
The response variable of each current build is determined based on whether the next build entry (\ie\ based on building date and time) for the same pull request has a \emph{changed review comment statistic} or not. 
Thus, the response variable for each of our build entries takes one of these two values--\emph{``new review"} or \emph{``unchanged."}
We used a randomly sampled set of 26,516 build entries from our dataset for the experiments where equal number of instances from both classes are ensured.
We use \emph{Naive Bayes (NB)}, \emph{Logistic Regression (LR)} and \emph{J48} from WEKA \cite{weka} workbench for the training, and apply 10-fold cross validation for the testing of our prediction models. 
Table \ref{table:result} summarizes our findings. 

\begin{table}
\centering
\caption{Performance of Prediction Models}\label{table:result}
\vspace{-.2cm}
\resizebox{3.5in}{!}{%
\begin{threeparttable}
\begin{tabular}{l|l|c|c|c}
\hline
\multirow{2}{*}{\textbf{Algorithm}} & \multirow{2}{*}{\textbf{Metrics}}&\textbf{Overall} & \multicolumn{2}{c}{\textbf{New Review Triggered}}\\
\hhline{~~~--}
& & \textbf{Accuracy}& \textbf{Precision} & \textbf{Recall}\\
\hline
\hline
Naive & \{all metrics\} &  58.03\% & 68.70\% & 29.50\%\\
\hhline{~----}
Bayes & \{build status, code review\} &  56.50\% & 78.60\% & 17.80\%\\
\hline
\hline
Logistic & \{all metrics\} &  60.56\% & 64.50\% & 47.00\% \\
\hhline{~----}
Regression &  \{build status, code review\} & 60.18\% & 64.30\% & 45.60\%\\
\hline
\hline
\multirow{2}{*}{J48} &  \{all metrics\} & \textbf{64.04}\% & \textbf{69.50}\% & \textbf{50.10}\%\\
\hhline{~----}
 & \{build status, code review\} &  62.64\% & 73.80\% & 39.20\%\\
\hline
\end{tabular}
\centering
\end{threeparttable}
}
\vspace{-.6cm}
\end{table}

From Table \ref{table:result}, we see that J48-based model performed the best in separating review triggering build entries from the rest.  
Our model classifies the build entries with 64\% overall accuracy which is promising as a proof of concept.
Besides, the model can identify the true-positives with about 70\% precision and 50\% recall which are also promising. 
Such findings clearly answer our third research question regarding automatic prediction on code review triggering--\textbf{RQ$_3$}.

\section{Discussion \& Conclusion}
\label{sec:conclusion}
In this paper, we report an exploratory study using 578K automated build entries from MSR challenge dataset, where we investigate 
the impact of continuous integration on code reviews.  
We explore two different aspects of continuous integration--\emph{automated build status} and \emph{automated build frequency} and two aspects of code review--\emph{review participation} and \emph{review quality}, and investigate how the former aspects might affect the later. 
We apply several statistical tests, and perform correlation and comparative analysis to answer our three research questions.
Our findings both confirm intuitive beliefs and reveal new meaningful information as follows:  

Automated build status has a notable impact on code review participation where passed builds having the maximum influence.   
Our analyses show that automated builds triggered new code reviews for 28\% of the 31,648 pull requests from our dataset. 
While \emph{passed} builds played the major role, \emph{errored} and \emph{failed} builds also brought about new code changes and thus, triggered further code reviews for 10\% of the pull requests.

Build frequency has a significant impact on the quality of code reviews in the open source projects. 
Our analyses suggest that code review comments are significantly higher for frequently built projects than that of rarely built projects. 
Frequently built projects often maintain a steady level of code review activities over the years, which is most probably missing from the rarely built projects according to our findings.

Experiments with the three prediction models show that most of our identified metrics-- build status, code change statistics, test change statistics and review comment statistics-- are effective in predicting
whether an automated build might trigger new code reviews or not.
Since the dataset was skewed, we conduct experiments with a randomly sampled subset that contains equal number of instances from both classes.
Our model provides a reasonable accuracy of 64\% with up to 70\% precision and 50\% recall.
Although peer code reviews are reported to be effective for quality improvement of software systems, they are generally done manually and are often time-consuming. Our model can offer automatic supports in the code reviews and software quality management by identifying the appropriate pull requests for code reviews using build logs.    

\textbf{Acknowledgement:} This research was supported in part by the Natural Sciences and Engineering Research Council of Canada (NSERC).     



\balance

\bibliographystyle{plainnat}
\setlength{\bibsep}{0pt plus 0.3ex}
\scriptsize
\bibliography{sigproc}  
%
%
\end{document}